\documentclass[a4paper,12pt]{article}
\usepackage{graphicx}
\usepackage{epsfig}
\usepackage{comment}
\usepackage{amsmath,amsthm,placeins}
\usepackage{xspace}
\usepackage[colorlinks=true,citecolor=blue]{hyperref}
\usepackage[acronym,nonumberlist,nogroupskip]{glossaries}

\newcommand{\ie}{{\it i.e.}\ }

\newcommand{\Ito}{It\^{o} }

\newcommand{\gest}{g_{\text{est}}}

\newcommand{\ave}[1]{\left\langle#1 \right\rangle}

\newcommand{\Ns}{S}

\newcommand{\gave}{g_{\ave{}}}

\newcommand{\elabel}[1]{\label{eq:#1}}
\newcommand{\eref}[1]{Eq.~(\ref{eq:#1})}

\newcommand{\Eref}[1]{Equation~(\ref{eq:#1})}
\newcommand{\seclabel}[1]{\label{sec:#1}}
\newcommand{\secref}[1]{Sec.~(\ref{sec:#1})}

\newcommand{\flabel}[1]{\label{fig:#1}}
\newcommand{\fref}[1]{Fig.~\ref{fig:#1}}
\newcommand{\Fref}[1]{Figure ~\ref{fig:#1}}

\newcommand{\be}{\begin{equation}}
\newcommand{\ee}{\end{equation}}
\newcommand{\bea}{\begin{eqnarray}}
\newcommand{\eea}{\end{eqnarray}}
\newcommand{\bc}{\begin{center}}
\newcommand{\ec}{\end{center}}

\newcommand{\Wi}{W^{(i)}}

\begin{document}
\title{The sum of log-normal variates in geometric Brownian motion}
\author{Ole Peters and Alexander Adamou\\
London Mathematical Laboratory\\
8 Margravine Gardens, London W6 8RH, UK}
\date{\today}

\maketitle

\begin{abstract}
Geometric Brownian motion (GBM) is a key model for representing self-reproducing entities. 
Self-reproduction 
may be considered the definition of life \cite{Morowitz1992}, and the dynamics it induces are of interest to those 
concerned with living systems from biology to economics. 
Trajectories of GBM are distributed according to the well-known log-normal density, broadening with time. 
However, in many applications, what's of interest is not a single trajectory but the sum, or 
average, of several trajectories. The distribution of these objects is more complicated. Here we 
show two different ways of finding their typical trajectories. We make use of an intriguing connection to 
spin glasses: the expected free energy of the random energy model is an average of log-normal 
variates. We make the mapping to GBM explicit and find that the free energy result gives 
qualitatively correct behavior for GBM trajectories. We then also compute the typical sum of
lognormal variates using \Ito calculus. This alternative route is in close quantitative agreement 
with numerical work.
\end{abstract}

\section{Introduction -- why study GBM?}
The basic phenomenon of self-reproduction occurs from the 
smallest chemical compound capable of building copies of itself to arguably the most complex 
known social system, namely the global economy. Not only is the economy made by humans 
who self-reproduce and whose cells self-reproduce but capitalism itself taps into the power of 
self-reproduction: capital means resources which can be deployed in order to 
generate resources, which can be deployed in order\dots. Realism usually requires models that include noise, to account for the multitude of effects not explicitly modelled. GBM is a simple, intuitive, and analytically tractable model of noisy multiplicative growth. 
A deep understanding of this basic model of self-reproduction is crucial and only in the last few decades have 
we achieved this. 

A quantity, $x$, of self-reproducing resources (such as biomass or capital) follows GBM if it evolves over time, $t$, according to the \Ito stochastic differential equation,
\be
dx=x(\mu dt+\sigma dW_t).
\elabel{GBM}
\ee
$dt$ denotes the infinitesimal time increment and $dW_t$ the infinitesimal increment in a Wiener process, which is a normal variate with $\ave{dW_t}=0$ and $\ave{dW_t dW_s}=\delta(t-s) dt$. $\mu$ and $\sigma$ are constant parameters called the drift and volatility. Put simply, relative changes in resources, $dx/x$, are assumed to be drawn independently from a stationary normal distribution at each time step. \eref{GBM} represents the continuous-time limit of this process. Its solution,
\be
x(t)=x(0)\exp\left[\left(\mu-\frac{\sigma^2}{2}\right) t+\sigma W(t)\right],
\elabel{GBM_sol}
\ee
yields exponential growth at a noisy rate. Hereafter, we shall assume $x(0)=1$ and neglect it, except where retaining it is illustrative. The distribution of $x(t)$ is a time-dependent log-normal,
\be
\ln(x(t)) \sim \mathcal{N}\left( \left(\mu-\frac{\sigma^2}{2}\right)t,\, \sigma^2 t \right),
\ee
whose mean, median, and variance grow (or decay) exponentially in time:
\bea
\ave{x(t)} &=& \exp(\mu t); \elabel{mean_x}\\
\text{median}(x(t)) &=& \exp[(\mu-\sigma^2/2)t]; \elabel{median_x}\\
\text{var}(x(t)) &=& \exp(2\mu t)[\exp(\sigma^2t)-1]. \elabel{var_x}
\eea

GBM is a standard model in finance for self-reproducing quantities such as stock prices. Since relative changes are modelled as normal variates, the central limit theorem means that GBM is an attractor process for a large class of multiplicative dynamics. Any quantity whose relative changes are random variables with finite mean and variance will behave 
like a GBM after a sufficiently long time. We work with GBM because it is standard and general. It exemplifies important qualitative and universal features
of multiplicative growth.

Specifically, we are interested in GBM as a model of economic resources, owned by some entity. We will think of $x$ as measured in currency units, such as dollars.

\subsection{Non-ergodicity of GBM}
The non-ergodicity of this growth process
manifests itself in an intriguing way as a difference between the growth of the expectation
value of $x$ and the growth of $x$ over time. Imagine a world where people's wealth 
follows \eref{GBM}. In such a world each person's wealth grows exponentially at rate
\be
g_t=\mu-\sigma^2/2
\elabel{gt}
\ee
with probability 1 if we observe the person's wealth for a long time.
The expectation value of each person's wealth grows exponentially at
\be
g_{\ave{}}=\mu.
\elabel{gave}
\ee
The expectation value, by definition, is the average over an ensemble of $N$ realizations of
$x$ in the limit $N\to\infty$. Important insights follow: the aggregate wealth in our model 
economy does not grow at the same rate as an individual's wealth \cite{AdamouPeters2016} (meaning that GDP may be a flawed reflection of national
economic well-being); inequality grows indefinitely \cite{BouchaudMezard2000,BermanPetersAdamou2017} (even without interactions between individuals); and 
pooling and sharing resources accelerates growth \cite{PetersAdamou2015a,Bouchaud2015}.

\subsection{PEAs in GBM -- a sketch\seclabel{sketch}}
The following question often emerges in applications of these results. \Eref{gt} and \eref{gave} are limiting cases ($t \to \infty$ and $N\to\infty$, respectively): what happens when time and population size are finite? In \cite{PetersKlein2013} we studied the ``partial ensemble average'' (PEA) of GBM, details of which are in \secref{prevwork}. The PEA is the sample mean of $N$ independent realisations of GBM:
\be
\ave{x(t)}_N \equiv \frac{1}{N} \sum_{i=1}^N x_i(t).
\ee
Here we sketch out some simple arguments about how this object depends on $N$ and $t$.

Considering \eref{gt} and \eref{gave} together, we expect the following tension:
\begin{enumerate}
\item[A)] for large $N$, the PEA should resemble the expectation value, $\exp(\mu t)$;
\item[B)] for long $t$, all trajectories should grow like $\exp[(\mu-\sigma^2/2)t]$.
\end{enumerate}
Situation A -- when a sample mean resembles the corresponding expectation value -- is known in statistical physics as ``self-averaging.'' A simple strategy for estimating when this occurs is to look at the relative variance of the PEA,
\be
R \equiv \frac{\text{var}(\ave{x(t)}_N)}{\ave{\ave{x(t)}_N}^2}.
\ee
To be explicit, here the $\ave{\cdot}$ and $\text{var}(\cdot)$ operators, 
without $N$ as a subscript, 
refer to the mean and variance over all possible PEAs. The PEAs themselves, taken over finite samples of size $N$, are denoted $\ave{\cdot}_N$. 
Using standard results for the mean and variance of sums of independent random variables and inserting the results in \eref{mean_x} and \eref{var_x}, we get
\be
R(N) = \frac{e^{\sigma^2 t}-1}{N}.
\ee
If $R \ll 1$, then the PEA will likely be close to its own expectation value, which is equal to the expectation value of the GBM. Thus, in terms of $N$ and $t$, $\ave{x(t)}_N\approx\ave{x(t)}$ when
\be
t < \frac{\ln N}{\sigma^2}.
\elabel{short_t}
\ee
This hand-waving tells us roughly when the large-sample (or, as we see from \eref{short_t}, short-time) self-averaging regime holds. A more careful estimate of the cross-over time in \eref{t_c} is a factor of 2 larger, but the scaling is identical.

For $t>\ln N/\sigma^2$, the growth rate of the PEA transitions from $\mu$ to its $t\to\infty$ limit of $\mu-\sigma^2/2$ (Situation B). 
Another way of viewing this is to think about what dominates the average. For early times in the process, all trajectories are close together, but as time goes by the distribution broadens exponentially. Because each trajectory contributes with the same weight to the PEA, after some time the PEA will be dominated by the maximum in the sample,
\be
\ave{x(t)}_N \approx \frac{1}{N}\max_{i=1}^N \{x_i(t)\},
\ee
as illustrated in \fref{trajectories}.
Self-averaging stops when even the ``luckiest'' trajectory is no longer close to the expectation value $\exp(\mu t)$. This is guaranteed to happen eventually because the probability for a trajectory to reach $\exp(\mu t)$ decreases towards zero as $t$ grows. Of course, this takes longer for larger samples, which have more chances to contain a lucky trajectory. 

\begin{figure}
\centering
\includegraphics[height=9.3cm]{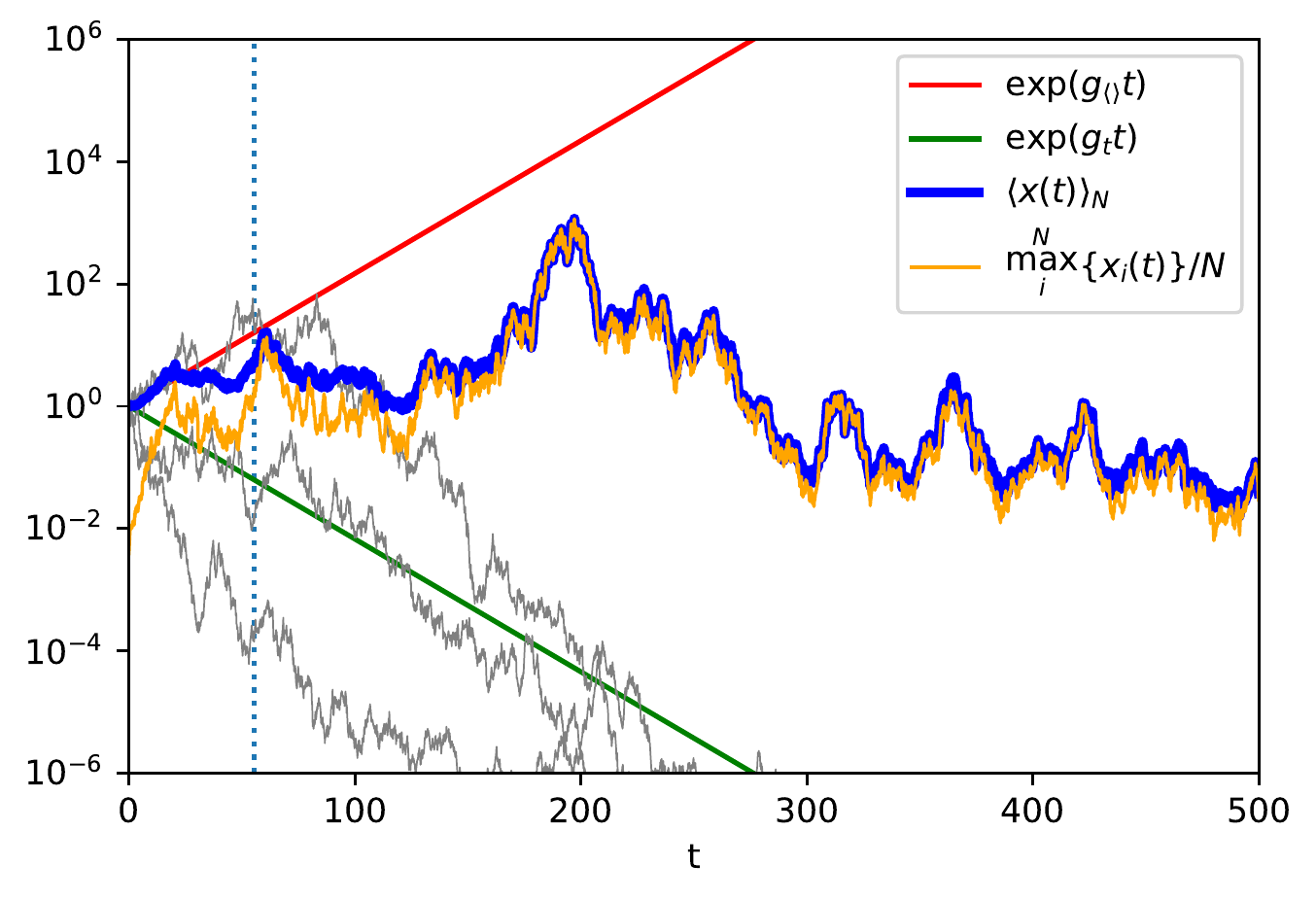}
\caption{PEA and maximum in a finite ensemble of size $N=256$. {\bf \underline{Red line:}} expectation value $\ave{x(t)}$. 
{\bf \underline{Green line:}} exponential growth at the time-average growth rate. In the $T\to\infty$ limit all trajectories grow at this rate. 
{\bf \underline{Yellow line:}} contribution of the maximum value of any trajectory at time $t$ to the PEA.  
{\bf \underline{Blue line:}} PEA $\ave{x(t)}_N$.
{\bf \underline{Vertical line:}} Crossover -- for $t>t_c=\frac{2\ln N}{\sigma^2}$ the maximum begins to dominate the PEA (the yellow line approaches the blue line).
{\bf \underline{Grey lines:}} randomly chosen trajectories -- any typical trajectory soon grows at the time-average growth rate.  
{\bf \underline{Parameters:}} $N=256$, $\mu=0.05$, $\sigma=\sqrt{0.2}$.}
\flabel{trajectories}
\end{figure}
\FloatBarrier

\subsection{Our previous work on PEAs\seclabel{prevwork}}
In \cite{PetersKlein2013} we analysed PEAs of GBM analytically and numerically. Using \eref{GBM_sol} the PEA can be written as
\be
\ave{x}_N=\frac{1}{N} \sum_{i=1}^N \exp\left[ \left(\mu-\frac{\sigma^2}{2}\right) t + \sigma \Wi(t) \right],
\elabel{PEA}
\ee
where $\left\{\Wi(t)\right\}_{i=1\dots N}$ are $N$ independent realisations of the Wiener process. Taking the deterministic part out of the sum we re-write \eref{PEA} as
\be
\ave{x}_N=\exp\left[ \left(\mu-\frac{\sigma^2}{2}\right) t \right] \frac{1}{N} \sum_{i=1}^N \exp\left(t^{1/2} \sigma \xi_i\right),
\elabel{PEA_2}
\ee
where $\left\{\xi_i\right\}_{i=1\dots N}$ are $N$ independent standard normal variates.

We found that typical trajectories of PEAs grow at $\gave$ up to a time $t_c$ 
that is logarithmic in $N$, meaning $t_c\propto \ln N$. This is consistent with our sketch in \secref{sketch}. After this time, typical 
PEA-trajectories begin to deviate from expectation-value behavior, and eventually 
their growth rate converges to $g_t$. While the two limiting behaviours $N\to\infty$
and $t\to \infty$ can be computed exactly, what happens in between
is less straight-forward. The PEA is a random object outside these limits. 
In \cite{PetersKlein2013} we dealt with this issue numerically by creating a super-sample
of $\Ns$ samples, each consisting of $N$ trajectories. In this way we were able to study the median of $\ave{x}_N(t)$, representing the behavior of typical trajectories.

A quantity of crucial interest to us is the exponential growth rate experienced by the PEA, 
\be
\gest(t,N) \equiv \frac{\ln(\ave{x(t)}_N)- \ln(x(0))}{t-0} = \frac{1}{t}\ln(\ave{x(t)}_N).
\elabel{gest}
\ee
In \cite{PetersKlein2013} we proved that the $t\to\infty$ limit for any (finite) 
$N$ is the same as for the case $N=1$, 
\be
\lim_{t\to\infty}\gest(t,N)=\mu-\frac{\sigma^2}{2}
\elabel{gest_2}
\ee
for all $N\geq1$. Substituting \eref{PEA_2} in \eref{gest} produces
\bea
\gest(t,N)&=&\mu-\frac{\sigma^2}{2}+\frac{1}{t} \ln\left(\frac{1}{N} \sum_{i=1}^N \exp( t^{1/2} \sigma \xi_i)\right)\\
&=&\mu-\frac{\sigma^2}{2}-\frac{\ln N}{t}+\frac{1}{t} \ln\left(\sum_{i=1}^N \exp( t^{1/2} \sigma \xi_i)\right).
\elabel{gest_4}
\eea

We didn't look in \cite{PetersKlein2013} at the expectation value of $\gest(t,N)$ for finite time and finite samples, but it's an interesting object that depends on $N$ and $t$ but is not stochastic. Note that this is not $\gest$ of the expectation value, 
which would be the $N\to\infty$ limit of \eref{gest}. Instead it is the 
$\Ns\to\infty$ limit,
\be
\ave{\gest(t,N)} = \frac{1}{t}\ave{\ln(\ave{x(t)}_N)} = f(N,t),
\elabel{gest_3}
\ee
where, as in \secref{sketch}, $\ave{\cdot}$ without subscript refers to the average over all possible samples, \ie $\lim_{\Ns\to\infty}\ave{\cdot}_{\Ns}$. The last two terms in \eref{gest_4} suggest an exponential relationship between ensemble size and time. The final term is a tricky stochastic object on which the properties of the expectation value in \eref{gest_3} will hinge. This term will be the focus of our attention: the sum of exponentials of normal random variates or, equivalently, log-normal variates.

\section{Mapping to the random energy model}
Since the publication of \cite{PetersKlein2013} we have learned, thanks to discussions with J.-P.~Bouchaud, 
that the key object in \eref{gest_4} -- the sum of exponentials of normal random variates -- has been of
interest to the mathematical physics community since the 1980s. 

The reason for this is Derrida's random energy model \cite{Derrida1980,Derrida1981}. It is defined as follows. 
Imagine a system whose energy levels are $2^K=N$ random numbers $\xi_i$ (corresponding to $K=\ln N/\ln 2$ spins). This is
a very simple model of a disordered system, such as a spin glass, the idea being that the system is so complicated
that we ``give up'' and just model its energy levels as realizations of a random variable. (We denote the number 
of spins by $K$ and the number of resulting energy levels by $N$, whereas Derrida uses $N$ for the number of spins).
The partition function is then
\be
Z=\sum_{i=1}^N \exp\left(\beta J\sqrt{\frac{K}{2}}\xi_i\right),
\elabel{Z}
\ee
where the inverse temperature, $\beta$, is measured in appropriate units, and the scaling in $K$ is chosen
so as to ensure an extensive thermodynamic limit \cite[p.~79]{Derrida1980}. $J$ is a constant that will be determined below.
The logarithm of the partition function gives the Helmholtz free energy, 
\bea
F&=&-\frac{\ln Z}{\beta}\\
&=&-\frac{1}{\beta}  \ln\left[\sum_{i=1}^N \exp\left(\beta J \sqrt{\frac{K}{2}}\xi_i\right)\right].
\elabel{F}
\eea

Like the growth rate estimator in \eref{gest}, this involves a sum of 
log-normal variates and, indeed, we can rewrite \eref{gest_4} as
\be
\gest=\mu-\frac{\sigma^2}{2}-\frac{\ln N}{t}-\frac{\beta F}{t},
\elabel{gest_5}
\ee
which is valid provided that
\be
\beta J \sqrt{\frac{K}{2}}=\sigma t^{1/2}.
\elabel{map}
\ee
\Eref{map} does not give a unique mapping between the parameters of our GBM, $(\sigma, t)$, and the parameters of the REM, $(\beta, K, J)$. Equating (up to multiplication) the constant parameters, $\sigma$ and $J$, in each model gives us a specific mapping:
\be
\sigma=\frac{J}{\sqrt{2}} \quad \text{and} \quad t^{1/2} = \beta\sqrt{K}.
\elabel{choice_1}
\ee

The expectation value of $\gest$ is interesting. The only random object
in \eref{gest_5} is $F$. Knowing $\ave{F}$ thus amounts to knowing $\ave{\gest}$.
In the statistical mechanics of the random energy model $\ave{F}$ is of key
interest and so much about it is known. We can use this knowledge
thanks to the mapping between the two problems.

Derrida identifies a critical temperature,
\be
\frac{1}{\beta_c} \equiv \frac{J}{2\sqrt{\ln 2}},
\elabel{beta_c}
\ee
above and below which the expected free energy scales differently with $K$ and $\beta$. This maps to a critical time scale in GBM,
\be
t_c = \frac{2\ln N}{\sigma^2},
\elabel{t_c}
\ee
with high temperature ($1/\beta>1/\beta_c$) corresponding to short time ($t<t_c$) and low temperature ($1/\beta<1/\beta_c$) corresponding to long time ($t>t_c$). Note that $t_c$ in \eref{t_c} scales identically with $N$ and $\sigma$ as the transition time, \eref{short_t}, in our sketch.

In \cite{Derrida1980}, $\ave{F}$ is computed in the high-temperature (short-time) regime as
\bea
\ave{F}&=&E-S/\beta \\
&=&-\frac{K}{\beta} \ln2 - \frac{\beta K J^2}{4},
\elabel{F_2}
\eea
and in the low-temperatures (long-time) regime as
\be
\ave{F}=-KJ\sqrt{\ln 2}.
\elabel{F_3}
\ee

\underline{Short time}\\
We look at the short-time behavior first (high $1/\beta$, \eref{F_2}).
The relevant computation of the entropy $S$ in \cite{Derrida1980} 
involves replacing the number of energy levels
$n(E)$ by its expectation value $\ave{n(E)}$. This is justified because
the standard deviation of this number is $\sqrt{n}$ and relatively small
when $\ave{n(E)}>1$, which is the interesting regime in Derrida's case. 

For spin glasses, the expectation value of $F$ is interesting, supposedly, 
because the system may be self-averaging and can be thought of as an
ensemble of many 
smaller sub-systems that are essentially independent. The macroscopic
behavior is then given by the expectation value.

Taking expectation values and substituting from \eref{F_2} in \eref{gest_5} we find
\be
\ave{\gest}^{\text{short}}=\mu-\frac{\sigma^2}{2}+\frac{1}{t} \frac{K J^2}{4T^2}.
\elabel{gest_6}
\ee
From \eref{map} we know that $t=\frac{KJ^2}{2\sigma^2T^2}$, which we substitute, to find
\be
\ave{\gest}^{\text{short}}=\mu,
\elabel{gest_7}
\ee
which is the correct behavior in the short-time regime.

\underline{Long time}\\
Next, we turn to the expression for the long-time regime (low temperature, \eref{F_3}). 
Again 
taking expectation values and substituting, this time from \eref{F_3} in \eref{gest_5}, we find
for long times
\be
\ave{\gest}^{\text{long}}=\mu-\frac{\sigma^2}{2}-\frac{\ln{N}}{t}+\sqrt{\frac{2\ln N}{t}}\,\sigma,
\elabel{gest_8}
\ee
which has the correct long-time asymptotic behavior.
The form of the correction to the time-average growth rate
in \eref{gest_8} is consistent with \cite{PetersKlein2013} and \cite{Redner1990}, where
it was found that approximately $N=\exp(t)$ systems are required for ensemble-average
behavior to be observed for a time $t$, so that the parameter $\ln N/t$ controls
which regime dominates -- if the parameter is small, then \eref{gest_8} indicates that the
long-time regime is relevant.

\Fref{1} is a direct comparison between the results derived
here, based on \cite{Derrida1980}, and numerical results using the same parameter 
values as in \cite{PetersKlein2013}, namely $\mu=0.05, \sigma=\sqrt{0.2}, N=256$ and $\Ns=10^5$.

Notice that $\ave{\gest}$ is not the (local)
time derivative $\frac{\partial}{\partial t}\ave{\ln(\ave{x}_N)}$, but a time-average growth rate, $\ave{\frac{1}{t}\ln\left( \frac{\ave{x(t)}_N}{\ave{x(0)}_N}\right)}$. 
In \cite{PetersKlein2013} we used a notation that we've stopped using since then because it
caused confusion -- $\ave{g}$ there denotes the growth rate of the expectation value, which 
is not the expectation value of the growth rate.

It is remarkable that the expectation value $\ave{\gest(N,t)}$ so closely reflects the
median, $q_{0.5}$, of $\ave{x}_N$, in the sense that
\be
q_{0.5}(\ave{x(t)}_N) \approx \exp \left(\ave{\gest(N,t)}t\right).
\elabel{quant_ave}
\ee
In \cite{PetersGell-Mann2016} it was discussed in detail that 
$\gest(1,t)$ is an ergodic observable for \eref{GBM}, in the sense that 
$\ave{\gest(1,t)}=\lim_{t\to\infty} \gest$. The relationship in \eref{quant_ave}
is far more subtle. The typical behavior of GBM PEAs 
is complicated outside the limits $N\to\infty$ or $t\to\infty$, in the sense that growth rates are 
time dependent here. This complicated behavior is well represented by an 
approximation that uses physical insights into spin glasses. Beautiful!

\begin{figure}
\centering
\includegraphics[height=9.3cm]{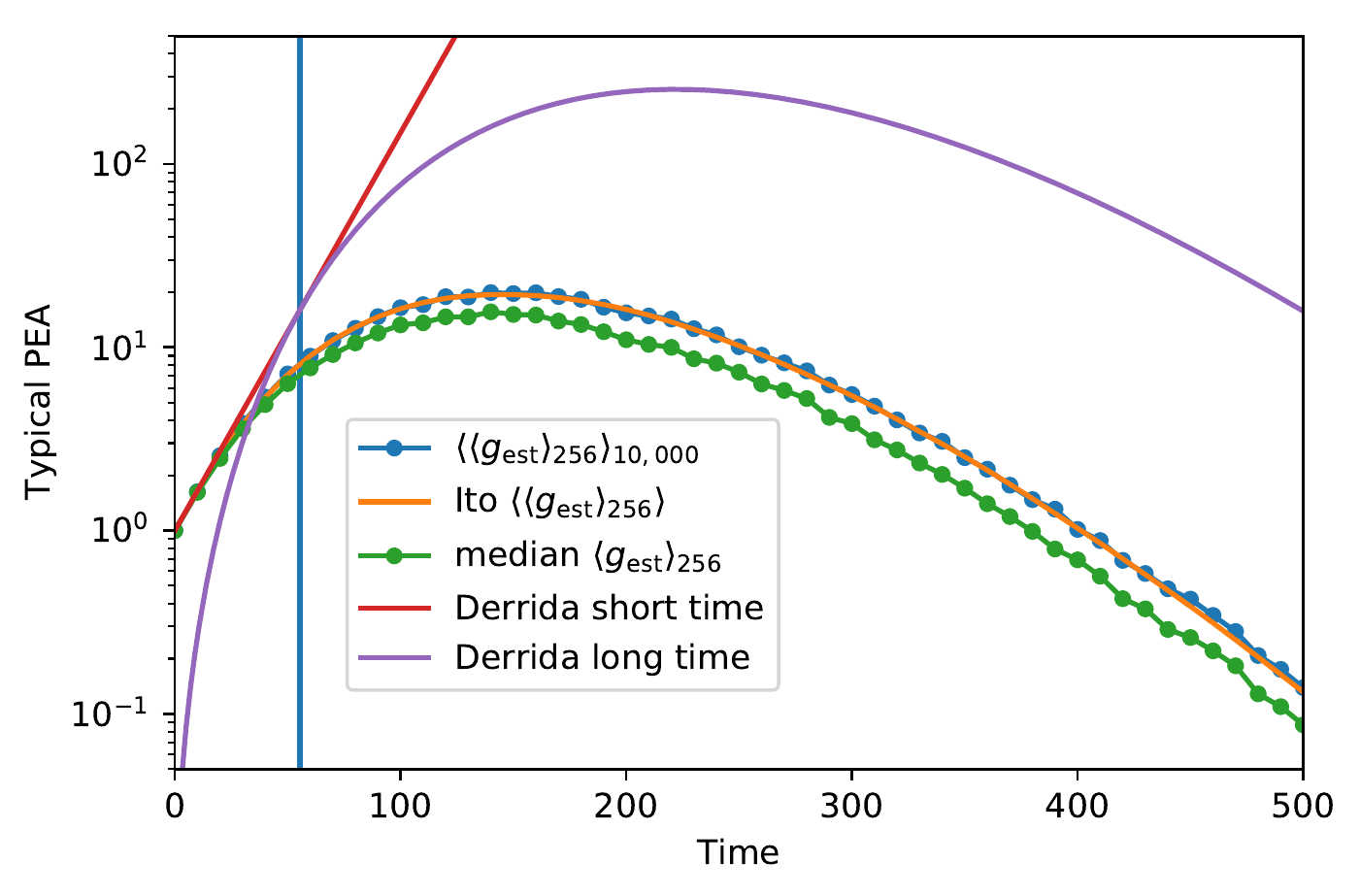}
\caption{Lines are obtained by exponentiating the various exponential 
growth rates. {\bf \underline{Blue line:}} $\ave{\ave{\gest}_{256}}_{10,000}$ is the numerical mean 
(approximation of the expectation value) 
over a super-ensemble of $\Ns=10,000$ samples of $\gest$ estimated in sub-ensembles of $N=256$ GBMs each. 
{\bf \underline{Green line:}} median in a super-ensemble of $\Ns$ samples of $\gest$, each estimated in sub-ensembles of size $N$. 
{\bf \underline{Yellow line:}} \eref{Ito_sums} is an exact expression for $d\ave{\ln\ave{x}_N}$, derived using \Ito calculus. We evaluate the expression by Monte Carlo, and integrate, $\ave{\ln\ave{x}_N}=\int_{0}^{t} d\ave{\ln\ave{x}_N}$. Exponentiation yields the yellow line. 
{\bf \underline{Red line:}} short-time behavior, based on the random energy model, \eref{gest_7}.
{\bf \underline{Purple line:}} long-time behavior, based on the random energy model, \eref{gest_8}. {\bf \underline{Vertical line:}} Crossover between the regimes at $t_c=\frac{2\ln N}{\sigma^2}$, corresponding to $\beta_c=\frac{2(\ln 2)^{1/2}}{J}$.
{\bf \underline{Parameters:}} $N=256$, $\Ns=10,000$, $\mu=0.05$, $\sigma=\sqrt{0.2}$.}
\flabel{1}
\end{figure}
\FloatBarrier

\section{Another route via \Ito calculus}
Another way to find the expectation value of PEA growth rates, \eref{gest_3}, 
is to compute $\ave{d\ln \ave{x}_N}$ using \Ito calculus. We compute this directly, without 
invoking the random energy model. To apply \Ito calculus, 
we will need the first two partial derivatives of $d\ln\ave{x}_N$ with respect to $x_i$.
\be
\frac{\partial \ln \ave{x}_N}{\partial x_i}=\frac{1}{N\ave{x}_N}
\ee
and
\be
\frac{\partial^2 \ln \ave{x}_N}{\partial x_i^2}=-\frac{1}{N^2\ave{x}_N^2}
\ee
Now, Taylor-expanding $d \ln \ave{x}_N$ we find
\bea
d \ln \ave{x}_N&=&\sum_i \frac{\partial \ln \ave{x}_N}{\partial x_i} dx_i+\frac{1}{2}\sum_i \sum_j \frac{\partial^2 \ln \ave{x}_N}{\partial x_i \partial x_j} dx_i dx_j+ \ldots\\
&\approx&\frac{1}{N\ave{x}_N} \sum_i dx_i-\frac{1}{2N^2\ave{x}_N^2}\sum_i \sum_j dx_i dx_j.
\eea
The double-sum can be split into `diagonal' ($i=j$) terms and cross-terms as
\be
\frac{1}{N\ave{x}_N}\sum_i x_i(\mu dt + \sigma dW_i) -\frac{1}{2N^2\ave{x}_N^2 }\left(\sum_i  dx_i^2 + \sum_{j}\sum_{i \neq j}dx_i dx_j\right)\hspace{.6cm}
\ee
Parts of the cross-terms are negligible because they are of order $dt^2$ and the rest vanishes when taking the expectation value, 
as we see by writing out one cross term
\be
dx_i dx_j=x_i x_j (\mu^2 dt^2+\mu\sigma dt dW_i+\mu \sigma dt dW_j + \sigma^2 dW_i dW_j).
\ee
We therefore drop these terms now, as we take the expectation value, using the Wiener identity $\ave{dW_i^2}=dt$ for the final term
\bea
\ave{d \ln \ave{x}_N}&=&\ave{\mu dt -\frac{1}{2N^2\ave{x}_N^2}\sum_i x_i^2 (\mu^2 dt^2 + \sigma^2 dt)} \\
&=&\mu dt   -\ave{\frac{1}{2} \sigma^2 \frac{\ave{x^2}_N}{N\ave{x}_N^2}} dt + O(dt^2).
\eea
Discarding terms of higher-than-first order in $dt$ and re-writing the last term as a fraction of sums, we thus have 
\bea
\frac{\ave{d \ln \ave{x}_N}}{dt}&=&\mu -\frac{1}{2} \sigma^2  \ave{\frac{\frac{1}{N}\sum_i x_i^2}{N\left(\frac{1}{N}\sum_i x_i\right)^2}}  \\
&=&\mu -\frac{1}{2} \sigma^2  \ave{\frac{\sum_i x_i^2}{\left(\sum_i x_i\right)^2}} \elabel{Ito_sums}
\eea
This expression has the correct behaviour: for short times, all $x_i$ are essentially identical, 
and the second term is $-\frac{1}{2}\sigma^2 \frac{1}{N}$, which is negligible if $N$ is large. So, for short times
we see expectation-value behaviour. For long times, the largest $x_i$ will dominate both the numerator and the denominator, and we have
$-\frac{1}{2}\sigma^2$: the full \Ito correction is felt for long times.

\section{Discussion}
The \Ito result is exact. A Monte-Carlo estimate of \eref{Ito_sums} (which is easy to obtain) is 
shown in \fref{1} (yellow line). This agrees well with numerical observations.
The approximations from the random energy model have the right shape and asymptotic behavior, 
though they're not on the same scale as the median PEA. This is, of course, not surprising because
these estimates are not designed to coincide with the median PEA. Quantitatively they are closer to 
a higher quantile of the distribution of PEAs. An intriguing question is this: is our computation 
using \Ito calculus helpful to compute the expected free energy of the random energy model? 

\bibliographystyle{abbrv}
\bibliography{bibliography}

\end{document}